\documentclass[floatfix,aps,amsmath,nofootinbib,onecolumn,11pt]{revtex4-1}
\def \be {\begin{equation}} 
\def \ee {\end{equation}} 
\def \bea {\begin{eqnarray}} 
\def \eea {\end{eqnarray}} 

\usepackage{graphicx}
\usepackage{dcolumn}
\usepackage{bm}
\usepackage{epsfig} 
\usepackage{amsfonts}
\usepackage{amsmath}
\usepackage{amssymb}
\usepackage[usenames]{color}
\usepackage[dvipsnames]{xcolor}
\usepackage[unicode, colorlinks=true, linkcolor=linkcolor, citecolor=linkcolor, filecolor=linkcolor,urlcolor=linkcolor, pdfusetitle]{hyperref}
\usepackage{natbib}
\usepackage{makecell}

\hypersetup{colorlinks,citecolor=blue,linkcolor=blue,urlcolor=blue}
\hypersetup{final=true}

\begin{document}

\title{Consistency tests between SDSS and DESI BAO measurements}

\author{Basundhara Ghosh}
\email{basundharag@iisc.ac.in}
\affiliation{Department of Physics, Indian Institute of Science, C.V. Raman Road, Bengaluru - 560012, Karnataka, India}

\author{Carlos Bengaly}
\email{carlosbengaly@on.br}
\affiliation{Observat\'orio Nacional, 20921-400, Rio de Janeiro - RJ, Brazil}

\date{\today}

\begin{abstract}

In this work, we investigate whether the baryon acoustic oscillation (BAO) measurements from redshift surveys, like the Sloan Digital Sky Survey (SDSS), and the Dark Energy Spectroscopic Instrument (DESI), are consistent with each other. We do so by obtaining the Hubble and deceleration parameter, respectively $H(z)$ and $q(z)$, from both datasets using a non-parametric reconstruction, so that our results do not depend on any {\it a priori} assumptions about the underlying cosmological model. We find that the reconstructed $H(z)$ and $q(z)$
from SDSS are significantly inconsistent with those obtained from DESI, and that both are only marginally consistent with the $\Lambda$CDM model ($\sim 3\sigma$ confidence level). Interestingly, the combined SDSS and DESI dataset reconciles with the standard model. These results are mostly unchanged with respect to different assumptions on the sound horizon scale value, as well as different reconstruction kernels. We also verify the results for the null diagnostic $\mathcal{O}_{\mathrm {m}}(z)$, finding that they suggest different trends for dark energy models using DESI and SDSS BAO measurements, and once again the combined dataset strongly agrees with $\Lambda$CDM. Therefore, our results call the attention for further examination of such inconsistency, as they can lead to biased and divergent results regarding the validity of the standard model, or the suggestion of new physics.  

\end{abstract}


\maketitle

\section{Introduction}\label{sec:intro}

The flat $\Lambda$CDM scenario has been established as the Standard Cosmological Model (SCM) for over two decades now. This model corresponds to a Universe dominated by cold dark matter, responsible for the formation of structures and dynamics of galaxies, and by the cosmological constant $\Lambda$ as the most successful candidate for the so-called dark energy, which accounts for the accelerated expansion of the Universe at late times.  Such a paradigm provides the best explanation, so far, of a variety of cosmological observations, e.g. the Cosmic Microwave Background (CMB) \cite{Planck:2018vyg}, the luminosity distances of Type Ia Supernovae (SNe) \cite{Brout:2022vxf,Rubin:2023ovl,DES:2024tys}, and galaxy clustering and weak lensing \cite{eBOSS:2020yzd,Heymans:2020gsg,DES:2021wwk}, thus validating the $\Lambda$CDM model as the SCM. Nevertheless, there are unresolved problems in relation to this model, such as the problems of primordial singularity and cosmic coincidence, besides tensions in measurements of some cosmological parameters, the most prominent being the tension of $\sim 5\sigma$ between the Hubble Constant $H_0$ measurements in the late- and early-time Universe with SNe and CMB, respectively~\cite{DiValentino:2021izs,Perivolaropoulos:2021jda}. These issues bring the validity of the $\Lambda$CDM model into question, and demand a further examination of cosmological models and observations to verify whether there is evidence for physics beyond the SCM, or perhaps unaccounted systematics. 

Recently, the baryon acoustic oscillation (BAO) measurements from the first data release of the Dark Energy Spectroscopic Instrument (DESI) showed that there could be an evolution of the dark energy equation of state across cosmic timescales. Such a result hints at a possible breakdown of the cosmological constant paradigm~\cite{DESI:2024mwx}, especially when combined with the Dark Energy Survey 5 Year SN compilation~\cite{DES:2024tys} and the Planck CMB priors~\cite{Planck:2018vyg}. Many subsequent works assessed this issue, where most of them showed that the $z=0.51$ and $z=0.71$ BAO data points could be responsible for this result, although a possible bias due to the choice of the dark energy parameter priors has also been pointed out~\cite{Dinda:2024ktd,Patel:2024odo, LHuillier:2024rmp,Orchard:2024bve,Liu:2024gfy,Chudaykin:2024gol,Notari:2024rti,Gialamas:2024lyw,DESI:2024ude,Jia:2024wix, Mukherjee:2024ryz, Dinda:2024kjf,DESI:2024aqx, Carloni:2024zpl,Colgain:2024xqj,Cortes:2024lgw,Luongo:2024fww,Dinda:2024ktd,Jiang:2024xnu,Wolf:2023uno,DESI:2024kob}. A dynamical dark energy equation of state offers intriguing possibilities for cosmology, particularly in addressing key cosmological tensions such as the Hubble tension and the $S_8$ tension. Many studies have explored how dynamical dark energy scenarios can incorporate rich physics, including phantom crossings and negative dark energy densities at high redshifts, which can be potential avenues for alleviating the Hubble tension \cite{DiValentino:2019dzu,Vagnozzi:2019ezj,Alestas:2020mvb,Tiwari:2023jle}. Notably, recent DESI results from the $w_0w_a$ parametrisation (also known as the Chevallier-Polarski-Linder or the CPL parametrisation) suggest a phantom regime at high redshifts. It would be valuable to include DESI data in the analysis of dynamical dark energy models aimed at resolving these tensions. However, a recent study indicates that DESI data may not support the resolution of the $H_0$ tension with evolving dark energy, adding complexity to this approach \cite{Wang:2024dka}. Nevertheless, such a result is definitely interesting and worth investigating further.

In this work, we focus on a comparison between the state-of-the-art BAO measurements from DESI and the previous releases of Sloan Digital Sky Survey (SDSS). Recently, a comparison between DESI and non-DESI data has been carried out \cite{Park:2024jns}, with the latter including CMB anisotropy data and non-CMB data, where it was found that such a data compilation is better constrained to favour a flat $w_0w_a$CDM model, compared to a flat $\Lambda$CDM model. Our goal, however, is to verify in a model-independent way whether there is any inconsistency between the DESI and SDSS BAO datasets, as it would indicate potential systematics that might lead to discrepant conclusions about the validity of the SCM, as well as the evidence of new physics. We perform this comparison by means of non-parametric reconstructions of observable quantities, such as the cosmic expansion rate through the Hubble parameter, $H(z)$, and the deceleration parameter, $q(z)$, that can be obtained from each individual dataset and their combination. These reconstructions allow us to circumvent {\it a priori} conclusions due to the assumption of a cosmological model to describe those quantities of interest. Other model-agnostic studies of dark energy in the light of DESI have been performed in \cite{Mukhopadhayay:2024zam,Reboucas:2024smm,Ruchika:2024lgi,Yang:2024kdo}, and some interesting works on obtaining model-independent distances from BAO can be found in \cite{Anselmi:2017cuq,ODwyer:2019rvi,Anselmi:2022exn}. Interestingly, we find that the DESI and SDSS BAO datasets are significantly inconsistent with each other, and only marginally consistent with the SCM predictions, although the combined SDSS and DESI dataset is in great agreement with $\Lambda$CDM. This result is further illustrated by the null diagnostic $\mathcal{O}_{\mathrm {m}}$, which shows discrepant deviations from the SCM scenario between each individual dataset, yet again the agreement with the SCM is restored when the combined dataset is considered. 

The paper is structured as follows: Section~\ref{sec:analysis} is devoted to the description of the observational data, the theoretical framework, and the methodology of our analysis; Section~\ref{sec:res} presents the results obtained from this analysis; Section~\ref{sec:conc} is dedicated to the discussion of our results and the concluding remarks. 

\section{Analysis}\label{sec:analysis}

\subsection{Theoretical framework}

The standard model of Cosmology is based upon two fundamental pillars: (i) the validity of the cosmological principle (CP), i.e., the assumption of statistical isotropy and homogeneity in the large-scale Universe; (ii) the validity of the general relativity (GR) as the underlying theory of gravity. We assume that both hypotheses hold in our analysis (for a broader discussion on potential deviations from them, see~\cite{Perivolaropoulos:2021jda}) since we are more focused on identifying possible inconsistencies between different datasets and their possible cosmological consequences, than directly exploring plausible models beyond the SCM. Such an exercise will be left for future work. 

In the framework of GR and CP, assuming spatial flatness and considering a general evolving dark energy model, the cosmic expansion rate is given by the Hubble parameter, which reads
\begin{eqnarray}\label{eq:hz_de}
\left[\frac{H(z)}{H_0}\right]^2 &=& \Omega_{\rm m}(1+z)^3 + \Omega_{\rm DE} \exp{\left[3\int_0^z \frac{1+w(z')}{1+z'}dz'\right]}\,,
\end{eqnarray}
where $H_0$ denotes the Hubble constant, $\Omega_{\rm m}$ the total matter density parameter, i.e., baryon plus cold dark matter, $\Omega_{\rm DE} = 1-\Omega_{\rm m}$ the dark energy density parameter, $w(z)$ the dark energy equation of the state, and $z$ the redshift. By assuming that dark energy corresponds to the Cosmological Constant $\Lambda$, we have $w(z) = -1$, thus Eq.~\eqref{eq:hz_de} reduces to
\begin{equation}\label{eq:hz_lcdm}
\left[\frac{H(z)}{H_0}\right]^2 = \Omega_{\rm m}(1+z)^3 + (1-\Omega_{\rm m})\,.
\end{equation} 

We can define the deceleration parameter as
\begin{equation}\label{eq:qz}
q(z) = -\frac{\ddot{a}}{aH} = (1+z)\frac{H'(z)}{H(z)}-1\,,
\end{equation}
where $H'(z) \equiv dH(z)/dz$, and $\ddot{a} \equiv d^2a(t)/dt^2$, $a(t) \equiv (1+z)$, being the scale factor of the Universe, in addition to the null diagnostic $\mathcal{O}_{\rm m}(z)$, which is based on a consistency relation for the SCM~\cite{Zunckel:2008ti} (see also~\cite{Sahni:2008xx, Mortsell:2008yu} for similar tests):
\begin{equation}\label{eq:omz}
\mathcal{O}_{\rm m}(z) \equiv \frac{E(z)^2-1}{(1+z)^3 - 1} = \Omega_{\rm m} \;\; \mbox{in flat $\Lambda$CDM}\,,
\end{equation}
where $E(z) \equiv H(z)/H_0$, so that 
\begin{equation}\label{eq:null_test_01}
\mathcal{O}_{\rm m}(z) \neq \Omega_{\rm m} \;\; \mbox{implies that SCM is ruled out.}
\end{equation}
We remark that the deceleration parameter $q(z)$ is a purely kinematic quantity by definition. Therefore, it does not rely on the assumption of GR, as well as any other gravity model, depending solely on the validity of the CP. Nonetheless, it can be connected to the SCM by replacing Eq.~\eqref{eq:hz_lcdm} and its derivative w.r.t the redshift in Eq.~\eqref{eq:qz}. On the other hand, the null diagnostic $\mathcal{O}_{\rm m}(z)$ is defined in the scenario of both CP and GR.

\subsection{Reconstruction method}

The quantities we need to reconstruct from the BAO datasets, in order to perform our comparison, are $H(z)$, $H'(z)$, $q(z)$, and $\mathcal{O}_{\rm m}(z)$, along with their respective uncertainties. Since our goal is to avoid {\it a priori} assumptions about the underlying cosmology, we adopt a non-parametric approach using the Gaussian Process (GP) method.
By definition, a GP consists of a distribution over functions, rather than over variables as in the case of a Gaussian distribution. So, we can reconstruct a function from data points without explicitly assuming a parametrisation that would describe its relationship. We use the well-known {\sc GaPP} (Gaussian Processes in Python) package throughout this work~\cite{Seikel:2012uu}\footnote{\url{https://github.com/astrobengaly/GaPP}} (see also~\cite{Shafieloo:2012ht}, and e.g.~\cite{Seikel:2012cs,Busti:2014dua,Gonzalez:2017fra,Bengaly:2019ibu,Bernardo:2021qhu,Mukherjee:2021ggf,Rodrigues:2021wyk,Oliveira:2023uid,Favale:2024sdq,Lemos:2024jbl,Wang:2024rxm} for a non-extensive list of cosmological applications using {\sc GaPP}) in order to obtain $H(z)$ and $H'(z)$ from the DESI and SDSS BAO data. 

For the sake of computing the uncertainties of the $H(z)$ and $H'(z)$ reconstructions, namely $\sigma_{H(z)}$ and $\sigma_{H'(z)}$, we take the values provided by the {\sc GaPP} code after optimising the GP hyperparameters, assuming the squared exponential kernel (unless stated otherwise) for 250 evenly spaced-out bins across the $0<z<2.5$ interval. As for the uncertainties on the deceleration parameter and the null diagnostic, respectively, we error-propagate $q(z)$ and $\mathcal{O}_{\rm m}(z)$, as in Eqs.~\eqref{eq:qz} and ~\eqref{eq:omz}, which yields
\begin{equation}\label{eq:err_qz}
\left[\frac{\sigma_{\rm q(z)}}{1+q(z)}\right]^2 =  \left[\frac{\sigma_{H(z)}}{H(z)}\right]^2 + \left[\frac{\sigma_{H'(z)}}{H'(z)}\right]^2 - \left[\frac{2\sigma_{H(z)H'(z)}}{H(z)H'(z)}\right] \,,
\end{equation}
\begin{equation}\label{eq:err_omz}
\sigma_{\mathcal{O}_{\rm m}(z)} = \left[\frac{2E(z)}{(1+z)^3 - 1}\right]\sigma_{E(z)} \,,
\end{equation}
where $\sigma^2_{E(z)} = (\sigma^2_{H(z)}/H_0^2) + (H^2(z)/H_0^4)\sigma^2_{H_0}$, as in~\cite{Seikel:2012cs}. Note that we assume the $H_0$ measurements by Planck CMB, or SH0ES SN observations, when convenient, as priors in the computation of $\mathcal{O}_{\rm m}(z)$ and its respective derivative, i.e., Eqs.~\eqref{eq:omz} and~\eqref{eq:err_omz}, respectively. 

\subsection{Observational data}

The SDSS and DESI BAO measurements are provided in terms of three ratios:

(i) The $D_{\rm V}(z)/r_{\rm d}$, i.e., an averaged measurement of the three-dimensional BAO mode. The radial and transverse BAO modes are not disentangled from each other in this case, so that this ratio reads
\begin{eqnarray}\label{eq:dv_bao}
D_{\rm V}(z)/r_{\rm d} = \frac{[z D^2_{\rm M}(z)D_{\rm A}(z)]^{1/3}}{r_{\rm d}} \,.    
\end{eqnarray}

(ii) The $D_{\rm M}(z)/r_{\rm d}$, consisting of a transverse BAO mode measurement, given by
\begin{eqnarray}\label{eq:dm_bao}
D_{\rm M}(z)/r_{\rm d} = \frac{D_{\rm A}(1+z)}{r_{\rm d}} \,;    
\end{eqnarray} 

(iii) The $D_{\rm H}(z)/r_{\rm d}$ ratio, which is a measurement of the radial BAO mode, defined as
\begin{eqnarray}\label{eq:dh_bao}
D_{\rm H}(z)/r_{\rm d} = \frac{c}{H(z)r_{\rm d}} \,,
\end{eqnarray}
where the $r_{\rm d}$ in Eqs.~\eqref{eq:dv_bao} to~\eqref{eq:dh_bao} represents the sound horizon scale at the baryon drag epoch. Here, we assume the CMB Planck measurement for this parameter, $r_{\rm d} = 147.05 \pm 0.30$ Mpc~\cite{Planck:2018vyg}, unless stated otherwise. 

In the case of ratios (i) and (ii), shown in Eqs.~\eqref{eq:dv_bao} and~\eqref{eq:dm_bao}, respectively, the BAO measurements are given in terms of $D_{\rm M}(z)$ and $D_{\rm A}(z)$. They correspond to the radial comoving and angular diameter distances, respectively, according to~\cite{Hogg:1999ad}
\begin{eqnarray}\label{eq:Dz}
D_{\rm M}(z) = c\int_0^z \frac{dz'}{H(z')} \,, \quad D_{\rm A}(z) = \frac{D_{\rm M}(z)}{(1+z)} \,.
\end{eqnarray}

Since these two ratios are related to the integral of $H(z)^{-1}$, we would need to compute the first and second derivatives of $D_{\rm A}(z)$ in order to obtain $H(z)$ and $H'(z)$, respectively -- e.g. $D_{\rm H}(z) \equiv c/H(z) = D_{\rm M}
'(z)$~\cite{Dinda:2024ktd}, implying that ratio (ii) is the derivative of ratio (iii) w.r.t. the redshift -- and that there are correlations between those two ratios. Thus, contrary to parametric analysis cases, where the employment of the full BAO dataset is important for a more robust parameter estimation and model fitting, this would make our analysis more computationally costly, and further degrade the uncertainties of the $H(z)$ and $H'(z)$ reconstructions that are necessary to obtain $q(z)$ and $\mathcal{O}_{\rm m}(z)$. For this reason, we only adopt BAO measurements given in terms of the ratio (iii) in this work. Moreover, we remark that there are no expected correlations between the individual $D_{\rm H}(z)/r_{\rm d}$ BAO data points, since they cover different, non-overlapping redshift intervals, as we shall see with more details further on.

So, we transform the $D_{\rm H}(z)/r_{\rm d}$ measurements into $H(z)$ following Eq.~\eqref{eq:dh_bao}, obtain $\sigma_{H(z)}$ by error-propagating the corresponding $D_{\rm H}(z)/r_{\rm d}$ uncertainties, and perform the GP reconstruction over these quantities as explained in the previous subsection. Note that the Planck uncertainty around $r_{\rm d}$ is also error-propagated here, although it has minimal impact on the final results due to its small value.

\begin{table}[!t]
\parbox{\linewidth}{
    \centering
    \begin{tabular}{|c|c|c|c|}
    \hline
    \thead{$z$} & \thead{$z_\mathrm{eff}$} & \thead{$D_{\rm H}(z)/r_{\rm d}$} & \thead{$\mathrm{Reference}$} \\
    \hline 
    $0.2<z<0.5$ & 0.38 & $25.00 \pm 0.76$ & \cite{BOSS:2016wmc} \\
    $0.4<z<0.6$ & 0.50 & $22.33 \pm 0.58$ & \cite{BOSS:2016wmc} \\
    $0.6<z<1.0$ & 0.70 & $19.33 \pm 0.53$ & \cite{eBOSS:2020lta} \\
    $0.8<z<2.2$ & 1.48 & $13.26 \pm 0.55$ & \cite{eBOSS:2020gbb} \\
    $z>1.77$ & 2.33 & $9.08 \pm 0.34$ & \cite{eBOSS:2020tmo} \\
    \hline
    \end{tabular}
    \caption{Baryon acoustic oscillation (BAO) measurements from SDSS collaboration. From left to right, the first column shows the redshift range of each measurement, the second column shows the effective redshift, the third column shows the $D_{\rm H}(z)/r_{\rm d}$ ratios in Mpc units, along with its $1\sigma$ uncertainty, and the fourth column shows references of each measurement.
    \label{tab: Table1}}
}
\end{table}
\vspace{0.5cm}
\begin{table}[!t]
\parbox{\linewidth}{
    \centering
    \begin{tabular}{|c|c|c|c|}
    \hline
    \thead{$z$} & \thead{$z_\mathrm{eff}$} & \thead{$D_{\rm H}(z)/r_{\rm d}$} & \thead{$\mathrm{Reference}$}\\
    \hline
    $0.4<z<0.6$ & 0.51 & $20.98 \pm 0.61$ & \cite{DESI:2024mwx} \\
    $0.6<z<0.8$ & 0.71 & $20.08 \pm 0.60$ & \cite{DESI:2024mwx}  \\
    $0.8<z<1.1$ & 0.93 & $17.88 \pm 0.35$ & \cite{DESI:2024mwx} \\
    $1.1<z<1.6$ & 1.32 & $13.82 \pm 0.42$ & \cite{DESI:2024mwx} \\
    $1.77<z<4.16$ & 2.33 & $8.52 \pm 0.17$ & \cite{DESI:2024mwx} \\
    \hline
    \end{tabular}
    \caption{Same as Table~\ref{tab: Table1}, but rather for the DESI DR1 BAO measurements, according to \cite{DESI:2024mwx}. 
    \label{tab: Table2}}
}
\end{table}

As for the observational datasets, we utilise the latest $D_{\rm H}(z)/r_{\rm d}$ BAO measurements provided by the SDSS and DESI surveys, as displayed in Table~\ref{tab: Table1} and~\ref{tab: Table2}, respectively. In Table~\ref{tab: Table1}, from top to bottom, the first two SDSS data points were obtained from the Baryon Oscillation Spectroscopic Survey (BOSS) Galaxy sample of the SDSS-III Data Release 12~\cite{BOSS:2016wmc}, while the third and fourth ones come from the extended Baryon Oscillation Spectroscopic Survey (eBOSS) samples of Luminous Red Galaxy (LRG)~\cite{eBOSS:2020lta} and quasar (QSO) samples~\cite{eBOSS:2020gbb} of the SDSS-IV Data Release 16. The last one corresponds to the Lyman-$\alpha$ forest-QSO measurement from the same survey~\cite{eBOSS:2020tmo}. In Table~\ref{tab: Table2}, again from top to bottom, the data points correspond to the DESI samples of LRG1, LRG2, LRG3+ELG1, ELG2, and Lyman-$\alpha$ forest-QSO, respectively~\cite{DESI:2024mwx}. 

In addition, we combine the SDSS and DESI BAO datasets by the same token of~\cite{DESI:2024mwx}, in order to avoid double counting of the data. We adopt the measurements that encompass a larger cosmic volume when there is overlap of redshift range and cosmic tracer between the two surveys. Thus, we use the two lowest redshift data points of SDSS (respectively, $z=0.38$ and $z=0.50$), neglecting the DESI $z=0.51$ one. At higher redshifts, we adopt the LRG2, LRG3+ELG1, and ELG2 measurements by DESI instead, neglecting those from SDSS except for the $z=1.48$ eBOSS QSO data point. As for the Lyman-$\alpha$ forest, we use the combined “DESI+SDSS” measurement reported in~\cite{DESI:2024mwx}, that is, $D_{\rm H}(z=2.33)/r_{\rm d} = 8.72 \pm 0.14$, instead of the two individual data points by each survey. This combination is hereafter referred as DESI+SDSS C1. For the sake of an additional analysis, we swap the $z=0.50$ SDSS data point for the $z=0.51$ DESI one, and the same goes for the $z=0.71$ DESI measurement, which is replaced by the $z=0.70$ SDSS instead. This specific combination is henceforth named DESI+SDSS C2. Finally, we also note that we are not combining those datasets with SNe, since our goal is to directly compare different BAO datasets with as minimal information from other cosmological observations as possible -- except for the sound horizon scale prior.


\section{Results}\label{sec:res}

\begin{figure}[t]
    \centering
    \includegraphics[width=0.49\linewidth, height=7.0cm]{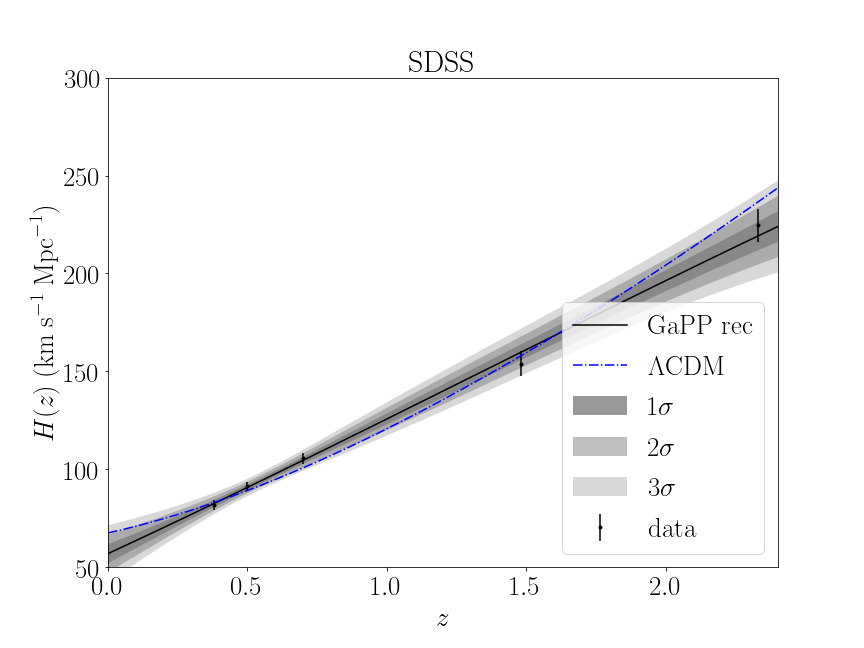}
    \includegraphics[width=0.49\linewidth, height=7.0cm]{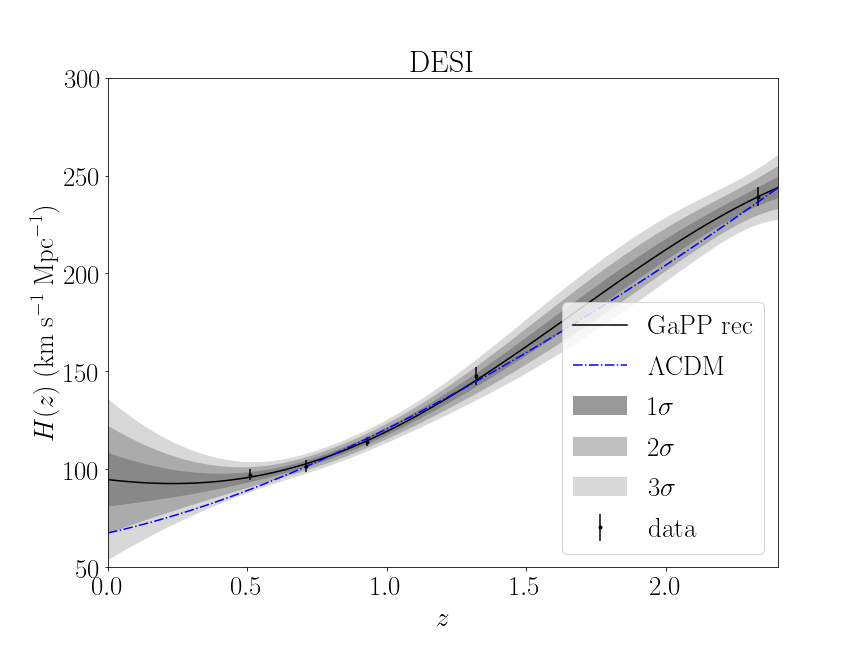}
    \includegraphics[width=0.49\linewidth, height=7.0cm]{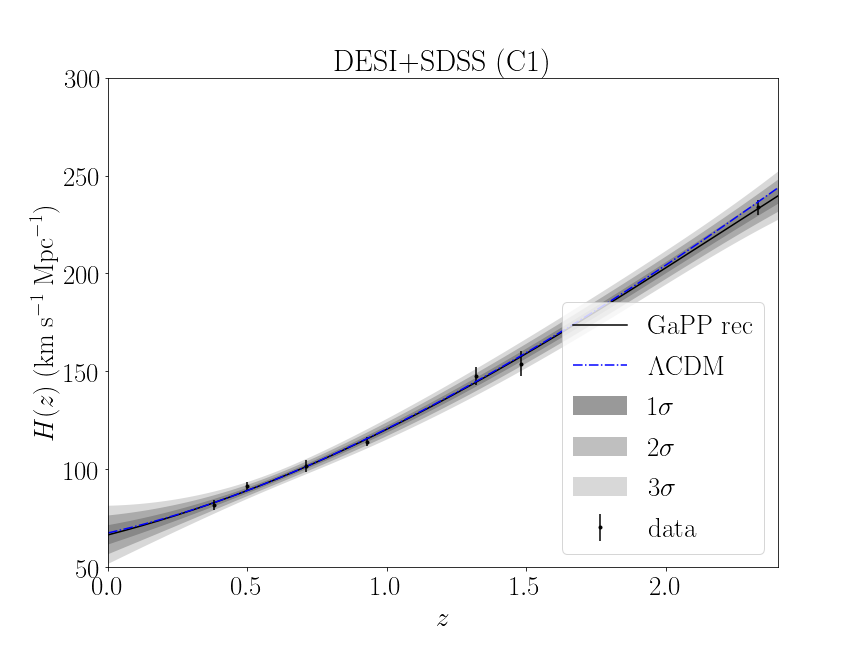}
    \includegraphics[width=0.49\linewidth, height=7.0cm]{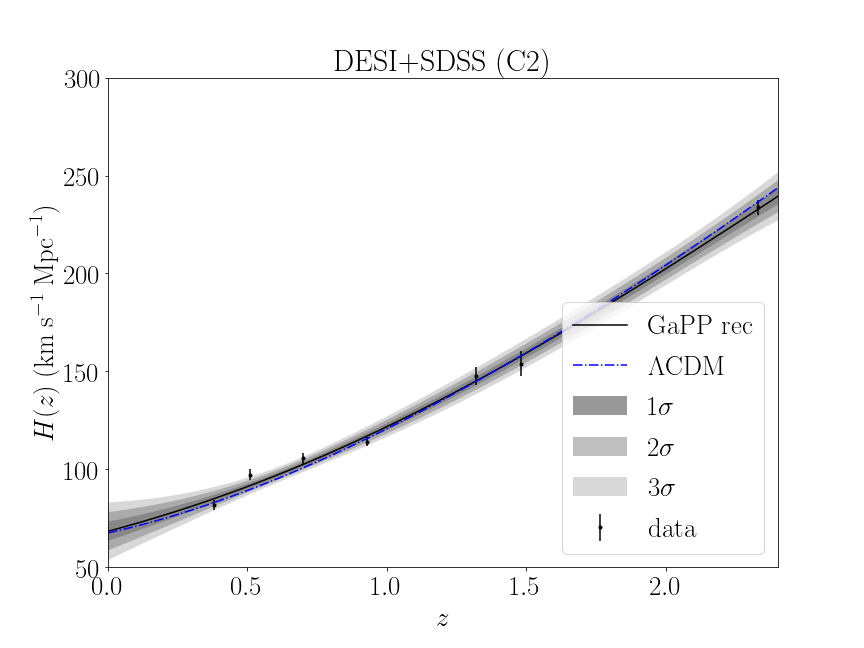}
    
    \caption{Reconstructed $H(z)$ for different datasets: SDSS (top left), DESI (top right), DESI and SDSS combination 1 (bottom left), and combination 2 (bottom right). The blue dot-dashed line denotes the SCM prediction, where we assume a flat $\Lambda$CDM model given by $\Omega_{\rm m} = 0.315$ and $H_0 = 67.4 \;\mathrm{km} \; \mathrm{s}^{-1} \; \mathrm{Mpc}^{-1}$, as reported by Planck 2018 results~\cite{Planck:2018vyg}. While the $\Lambda$CDM curve lies in the 3$\sigma$ band for both SDSS and DESI considered individually, the latter has a much larger uncertainty at low redshifts. In contrast, the combined data sets of SDSS and DESI exhibit a much better consistency with $\Lambda$CDM.}
    \label{fig:Hz_recon}
\end{figure}

\begin{figure}[t]
    \centering
    \includegraphics[width=0.49\linewidth, height=7.0cm]{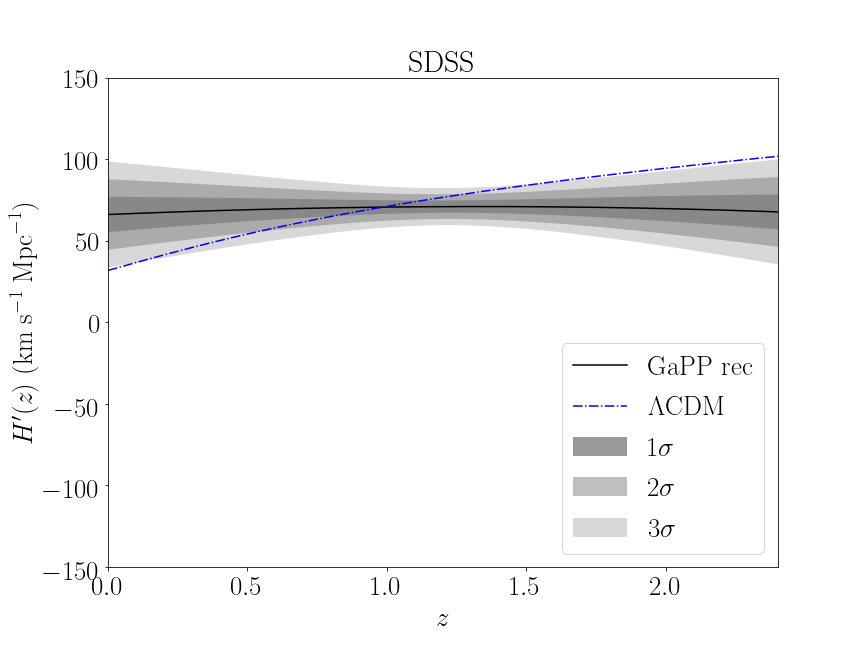}
    \includegraphics[width=0.49\linewidth, height=7.0cm]{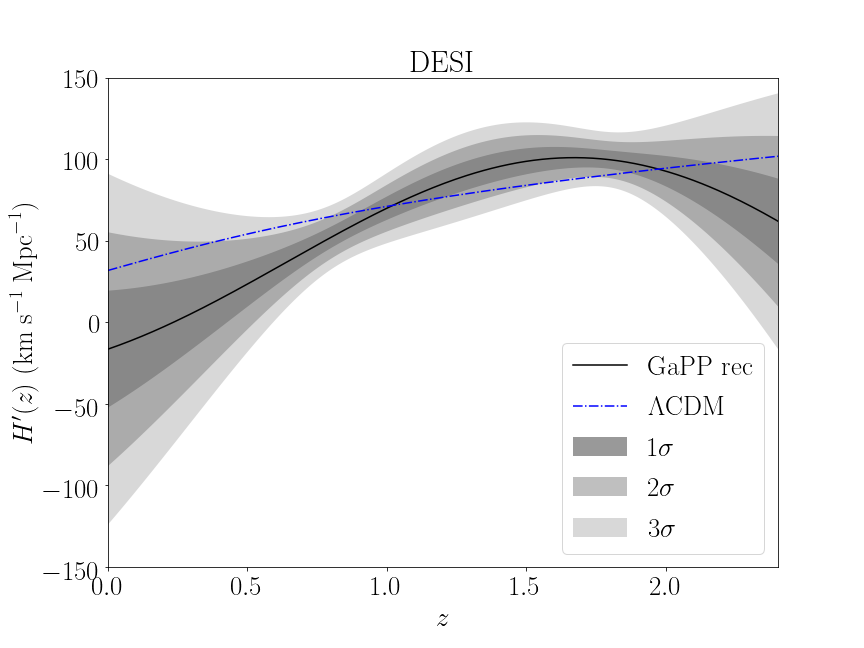}
    \includegraphics[width=0.49\linewidth, height=7.0cm]{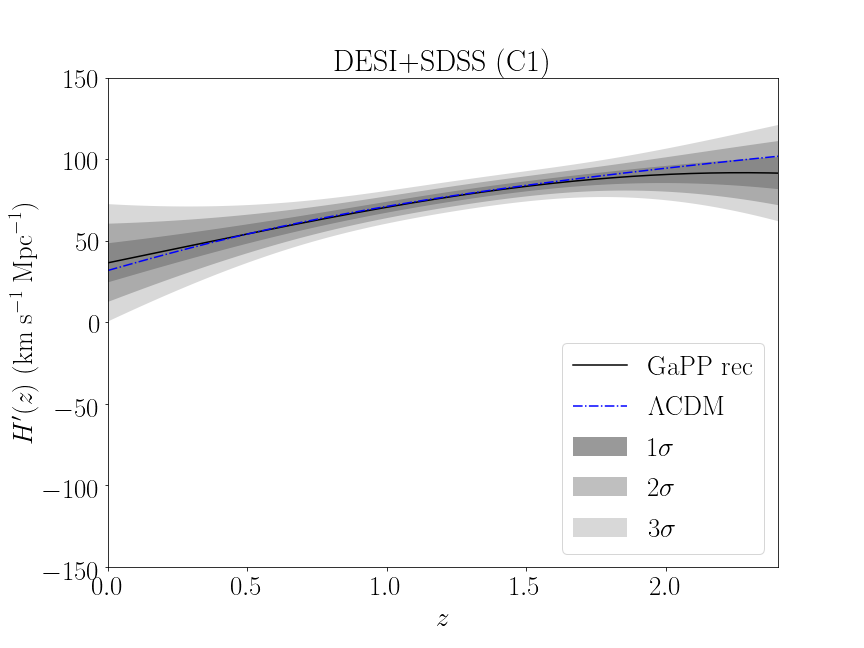}
    \includegraphics[width=0.49\linewidth, height=7.0cm]{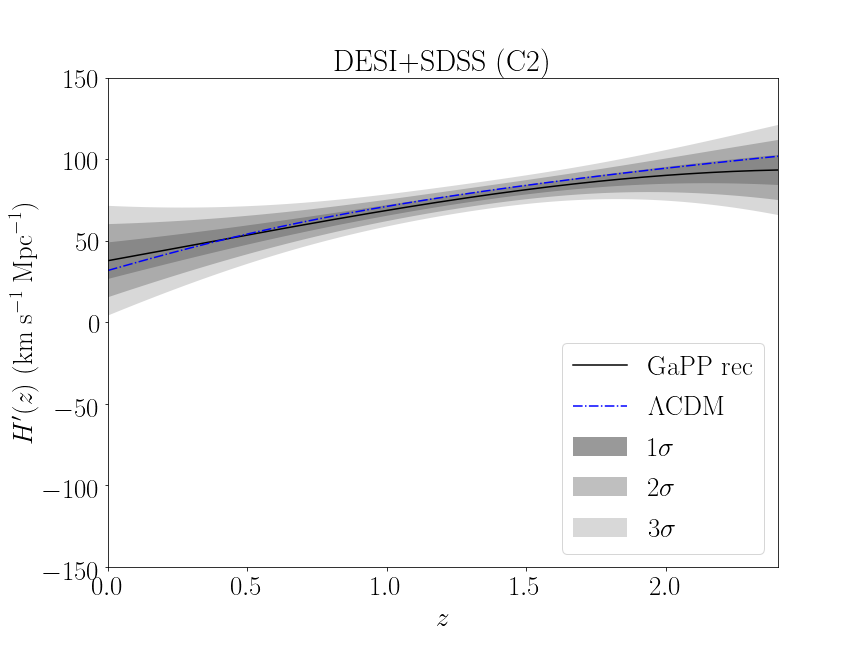}
    
    \caption{Same as Fig.~\ref{fig:Hz_recon}, but rather for its derivative $H'(z)$}\label{fig:dHz_recon}
\end{figure}

\begin{figure}[!t]
    \centering
    \includegraphics[width=0.49\linewidth, height=7.0cm]{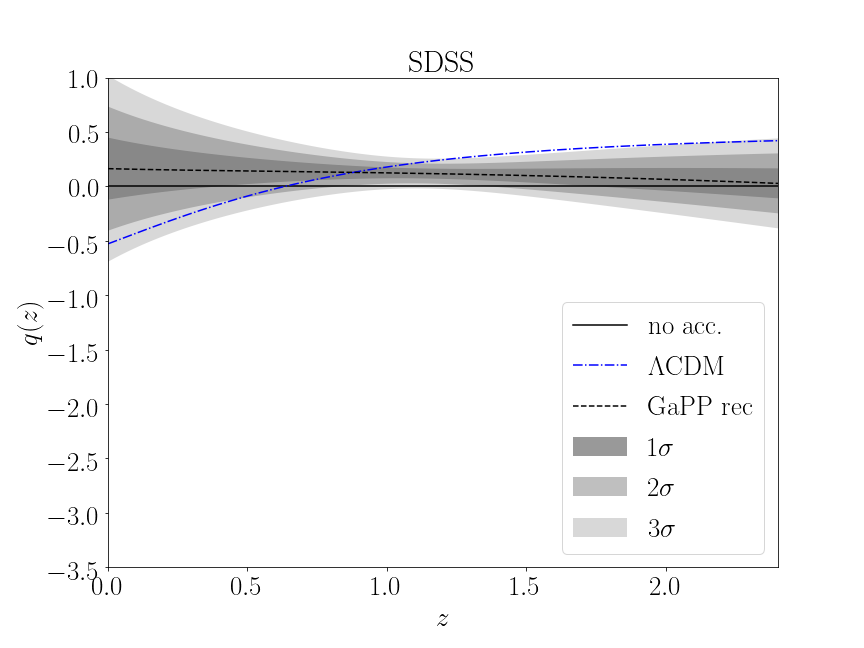}
    \includegraphics[width=0.49\linewidth, height=7.0cm]{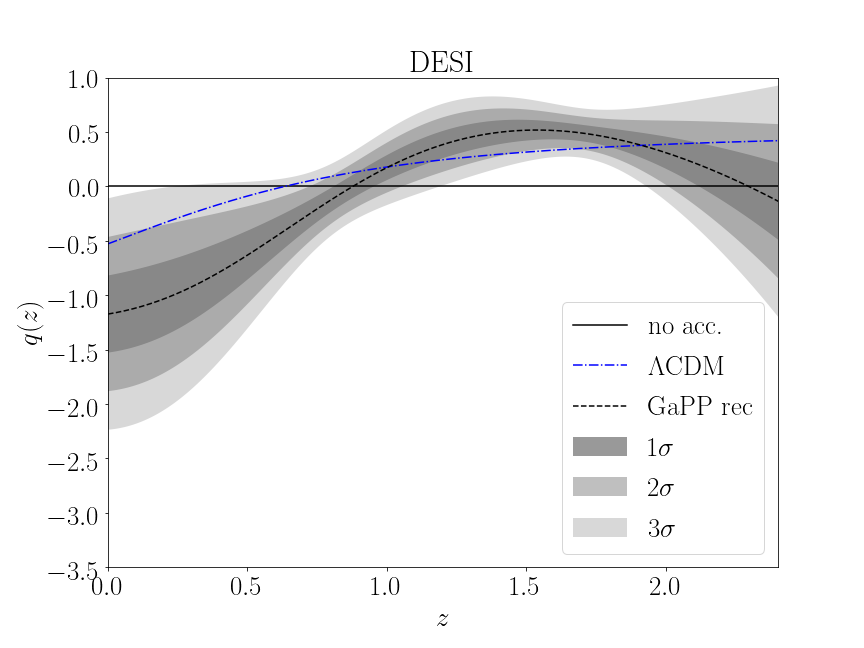}
    \includegraphics[width=0.49\linewidth, height=7.0cm]{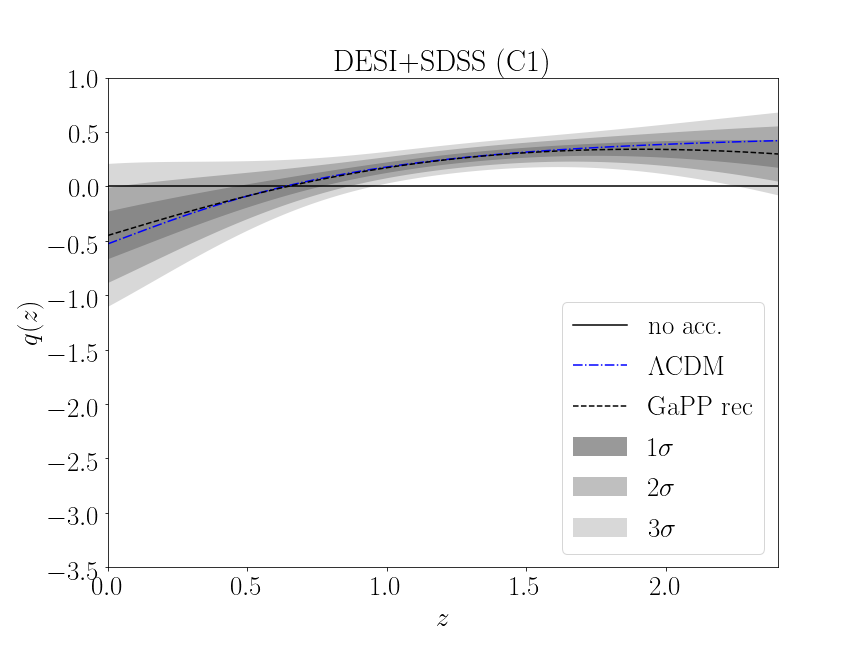}
    \includegraphics[width=0.49\linewidth, height=7.0cm]{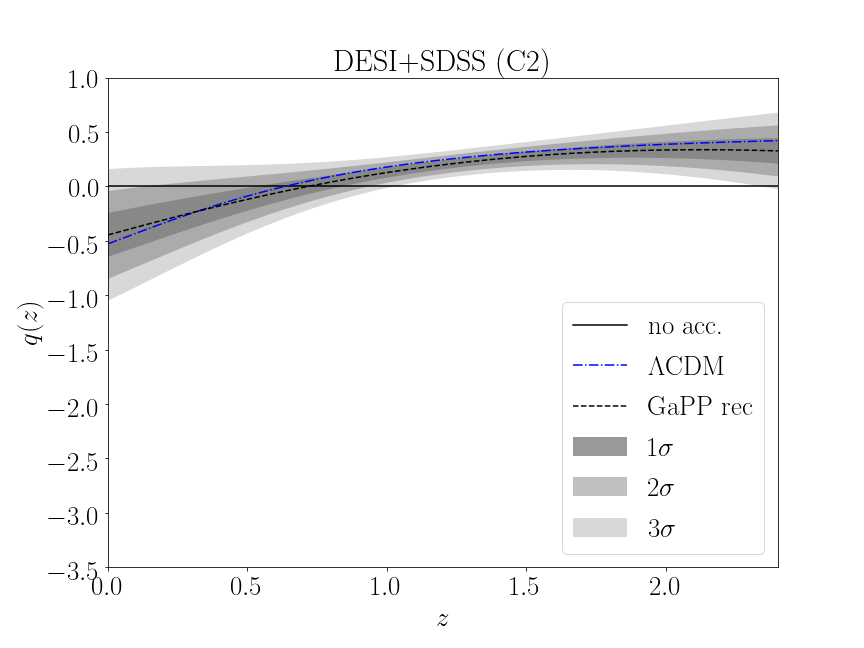}

    \caption{Same as Fig.~\ref{fig:dHz_recon}, but rather for the deceleration parameter $q(z)$}\label{fig:qz_recon}
\end{figure}

\begin{figure}[!t]
    \centering
    \includegraphics[width=0.49\linewidth, height=7.0cm]{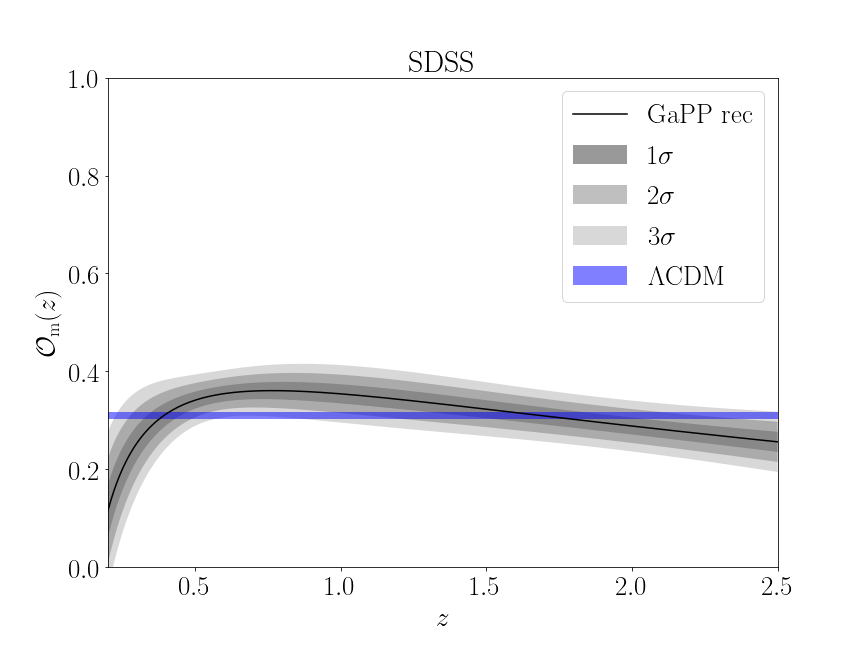}
    \includegraphics[width=0.49\linewidth, height=7.0cm]{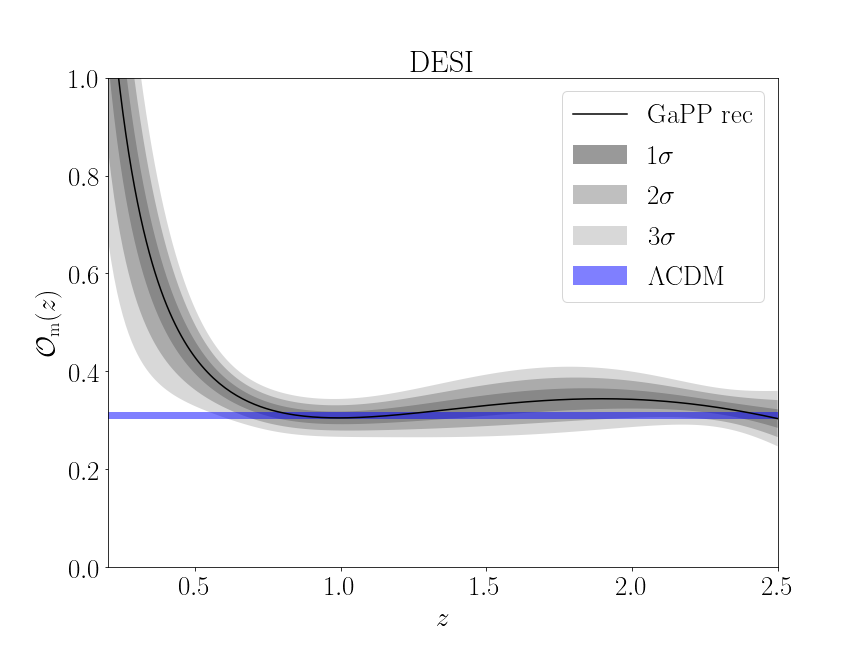}
    \includegraphics[width=0.49\linewidth, height=7.0cm]{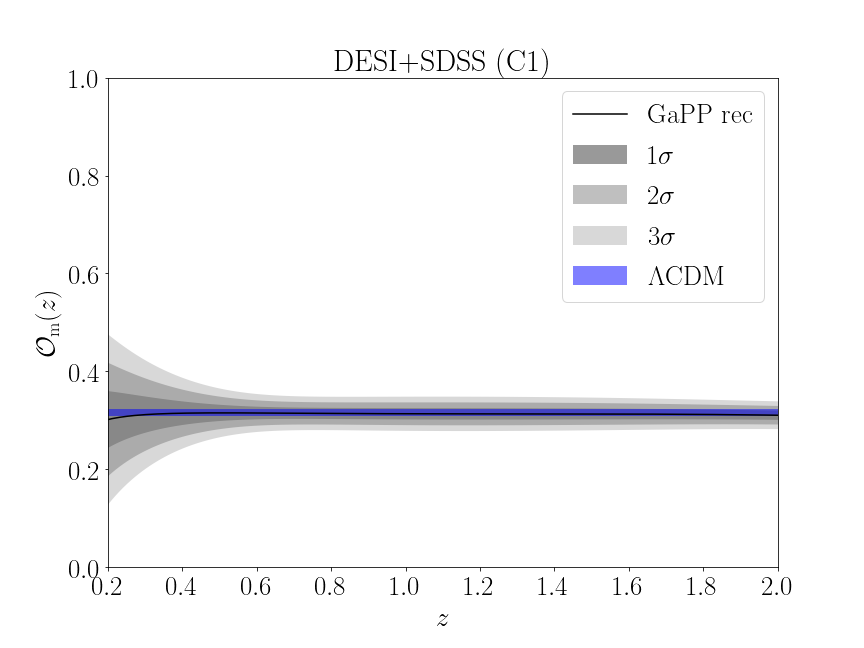}
    \includegraphics[width=0.49\linewidth, height=7.0cm]{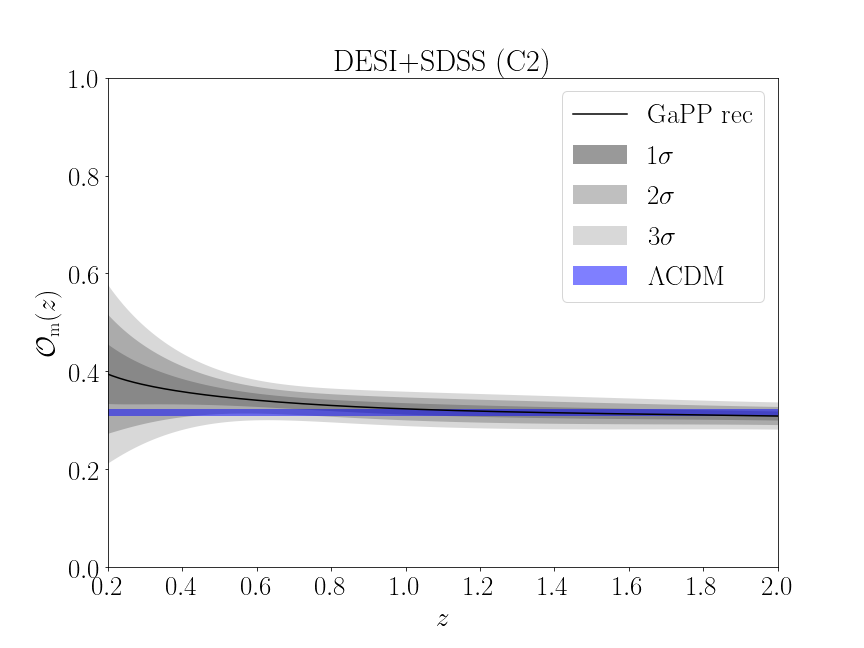}
    \caption{Same as Fig.~\ref{fig:qz_recon}, but rather for the null diagnostic $\mathcal{O}_{\rm m}(z)$. The thick blue line represents the latest Planck CMB (TT,TE,EE+lowE+lensing) constraint on $\Omega_{\rm m}$ at $1\sigma$ confidence level, that is, $\Omega_{\rm m} = 0.315 \pm 0.007$. We also assume the $H_0$ prior as the Planck 2018 best-fit in this case, i.e., $H_0 = 67.4 \pm 0.5 \; \mathrm{km} \; \mathrm{s}^{-1} \; \mathrm{Mpc}^{-1}$.}
    \label{fig:omz_recon}
\end{figure}

In this section, we present our reconstructed results and compare them with SCM predictions. Fig.~\ref{fig:Hz_recon} shows the reconstructed Hubble parameter as a function of redshift, $H(z)$ for the SDSS (upper left panel) and DESI (upper right panel) datasets alone, as well as combined together (lower panels). The blue dot-dashed lines represent the theoretical predictions for the SCM, where we assume a flat $\Lambda$CDM cosmology with $\Omega_{\rm m} = 0.315$ and $H_0 = 67.4 \; \mathrm{km} \; \mathrm{s}^{-1} \; \mathrm{Mpc}^{-1}$, as reported by Planck 2018 results~\cite{Planck:2018vyg}. 

As evident, there is a significant deviation from SCM at lower redshifts for both SDSS and DESI, with considerably higher uncertainties in case of DESI, which can be attributed to the lack of observational data at lower redshifts from the recent DESI results. In both cases, the SCM prediction curve lies in the 3$\sigma$ band, suggesting the possibility of beyond-SCM physics, but not ruling out SCM completely. The result obtained from DESI+SDSS C1 (bottom left panel), however, agree with SCM to a reasonable degree and lie in the 1$\sigma$ band. A similar result occurs for DESI+SDSS C2, as shown in the bottom right panel of the same figure. The only marginal difference between these reconstructions is the slightly larger uncertainties in the latter combination at lower redshifts, compared to the former case, because of the absence of the two data points mentioned. Nevertheless, the concordance with the SCM prediction is maintained.

Fig.~\ref{fig:dHz_recon} shows the reconstruction of the first derivative of the Hubble parameter, $H'(z)$, which agrees with the behaviour exhibited by the plots in Fig.~\ref{fig:Hz_recon}. The SDSS and DESI plots alone demonstrate a deviation from SCM, while the combined plots in the bottom panels are in good agreement with SCM predictions.

The reconstruction of the deceleration parameter $q(z)$ follows according to Eq.~\eqref{eq:qz} and \eqref{eq:err_qz}, and is shown in Fig.~\ref{fig:qz_recon}. Again, the top left panel display the reconstruction obtained from SDSS observations, the top right panel corresponds to the DESI case, while the bottom left and right panels, respectively, stand for the DESI+SDSS C1 and C2. The results are very curious, especially for SDSS only, where the {\sc GaPP} reconstruction seems to imply that the universe has never undergone an accelerated explanation. The reconstruction from DESI data, on the other hand, hints at an acceleration that might be slowing down at present times, consistent with the results put forth by the DESI collaboration -- but in stark contrast with our SDSS reconstruction. Interestingly, both combinations of the DESI and SDSS datasets give us a deceleration parameter that is in remarkable agreement with the SCM prediction. 

Similarly, in Fig.~\ref{fig:omz_recon}, the reconstruction of $\mathcal{O}_{\rm m}(z)$ (see Eq.~\eqref{eq:omz}) suggests a reasonable agreement with SCM when the SDSS and DESI datasets are combined. Note that the agreement is the best, that is within 1$\sigma$, for DESI+SDSS C1, whereas for DESI+SDSS C2 the SCM prediction lies in the 2$\sigma$ band of the {\sc GaPP} reconstruction. However, we find that the $\mathcal{O}_{\rm m}(z)$ results for DESI (SDSS) data alone suggest possible deviations from the standard model at over $3\sigma$ ($2\sigma$) confidence level, with different slopes for both cases, thus indicating different trends for dark energy.

\begin{figure}[!t]
    \centering
    \includegraphics[width=0.49\linewidth, height=7.0cm]{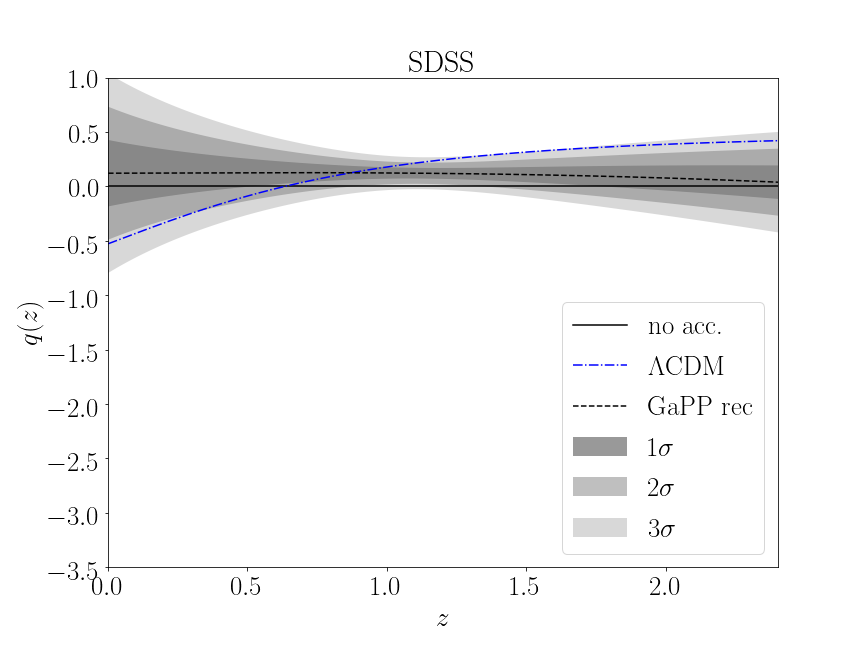}
    \includegraphics[width=0.49\linewidth, height=7.0cm]{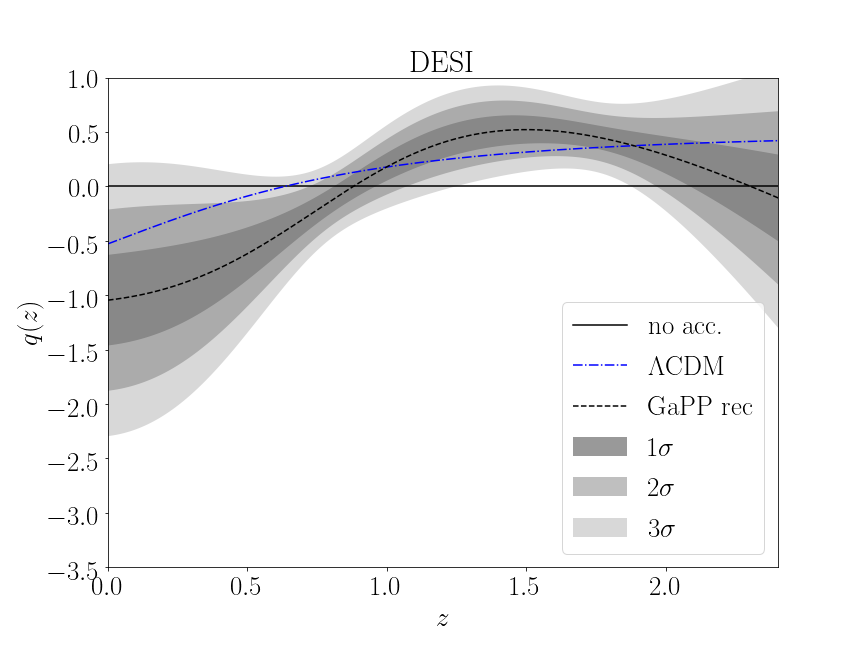}
    \includegraphics[width=0.49\linewidth, height=7.0cm]{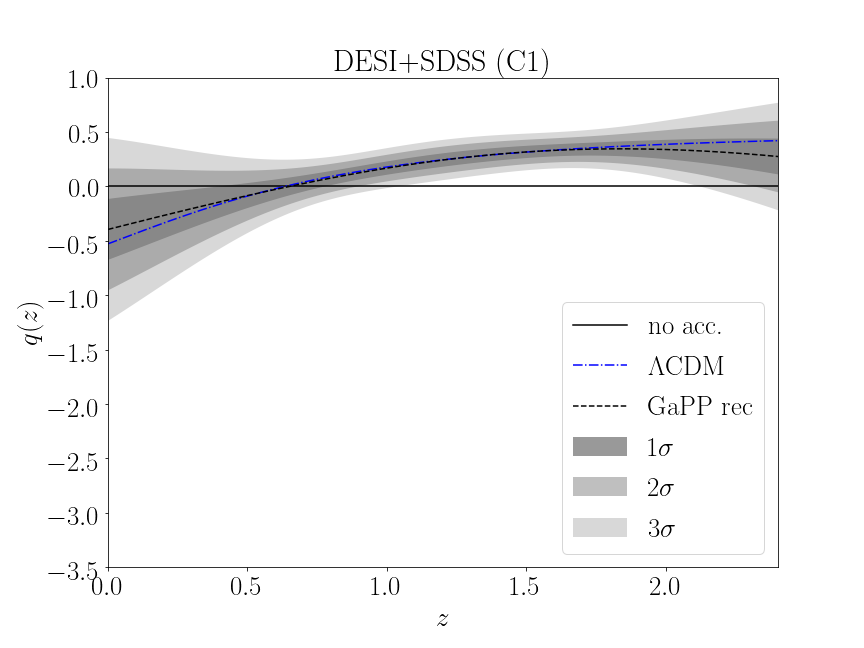}
    \includegraphics[width=0.49\linewidth, height=7.0cm]{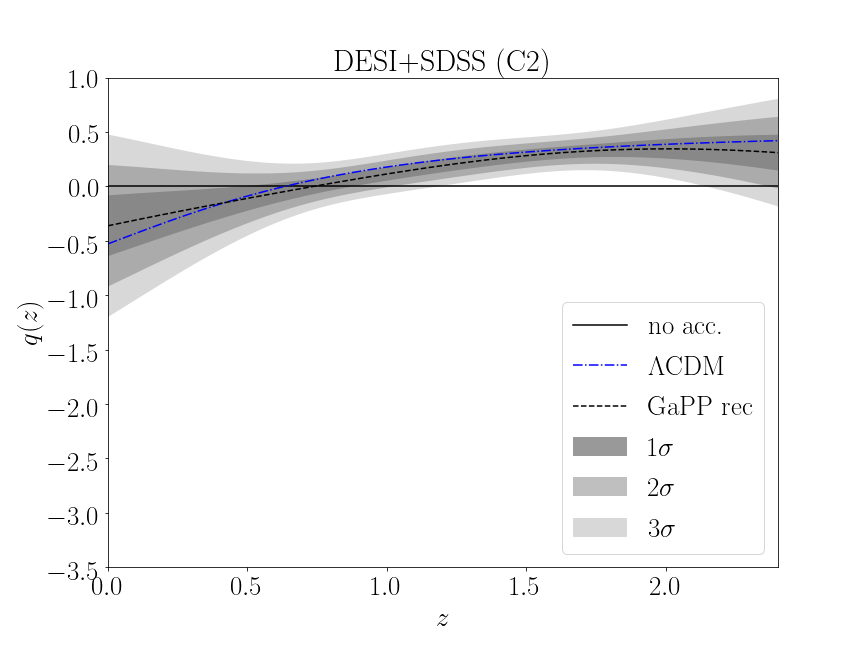}

    \caption{Same as Fig.~\ref{fig:qz_recon}, but assuming the Matern(7/2) GP kernel instead of the squared exponential one.}\label{fig:qz_mat72}
\end{figure}

\begin{figure}[!ht]
    \centering
    \includegraphics[width=0.49\linewidth, height=7.0cm]{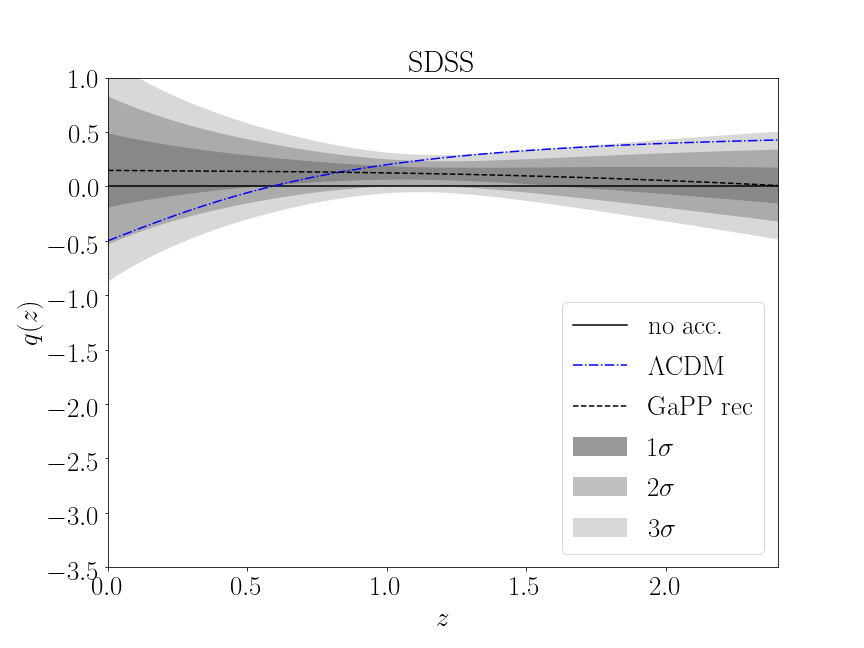}
    \includegraphics[width=0.49\linewidth, height=7.0cm]{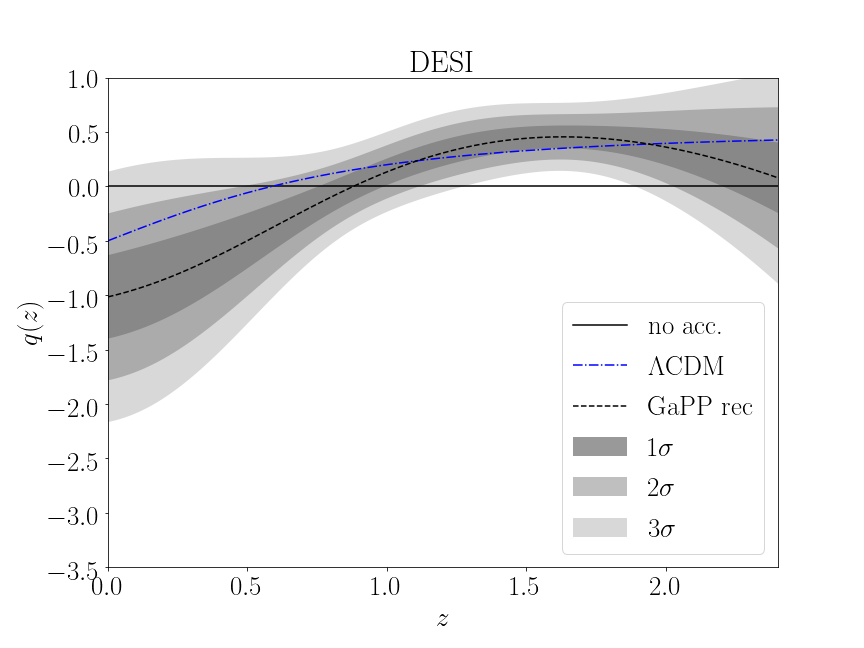}
    \includegraphics[width=0.49\linewidth, height=7.0cm]{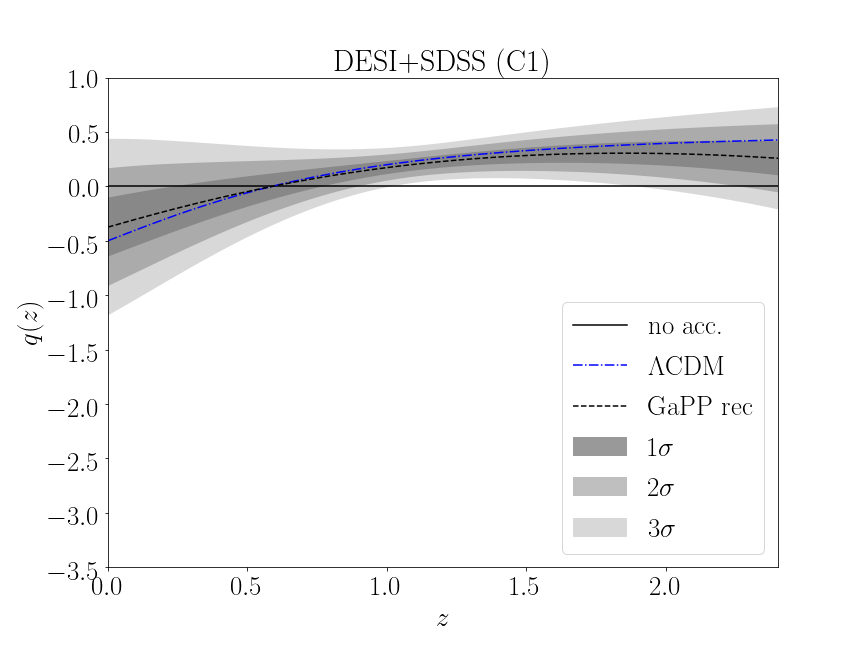}
    \includegraphics[width=0.49\linewidth, height=7.0cm]{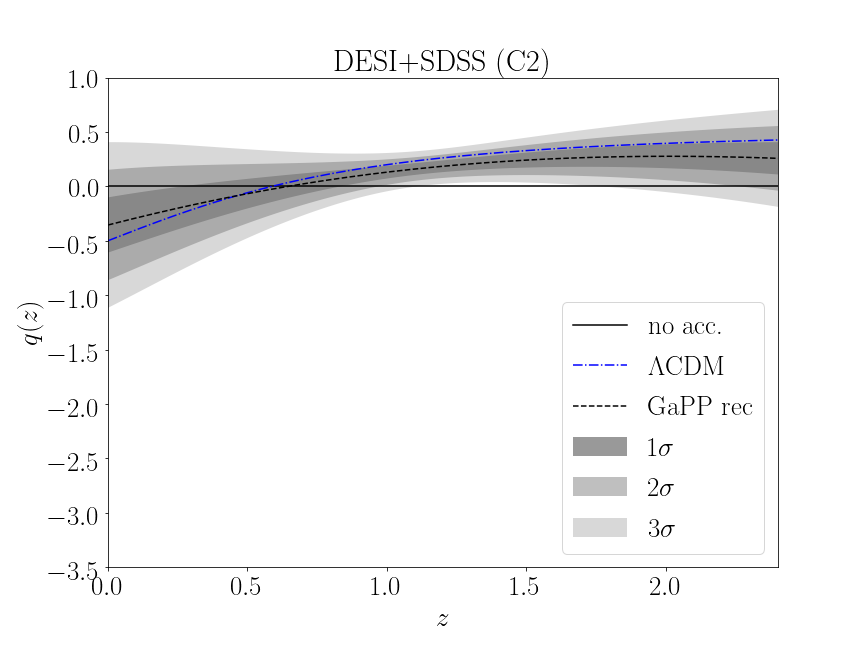}

    \caption{Same as Fig.~\ref{fig:qz_recon}, but assuming the low-$z$ prior on the sound horizon scale $r_{\rm d}$, such as $r_{\rm d} = 136.4 \pm 3.5$ Mpc, as shown in~\cite{Planck:2018vyg}. The blue dot-dashed lines correspond to a flat $\Lambda$CDM model given by $\Omega_{\rm m} = 0.334$ and $H_0 = 73.3 \; \mathrm{km} \; \mathrm{s}^{-1} \; \mathrm{Mpc}^{-1}$, which is consistent with the PantheonPlus and SH0ES latest results~\cite{Brout:2022vxf}.}\label{fig:qz_rdlowz}
\end{figure}

\begin{figure}[!t]
    \centering
    \includegraphics[width=0.49\linewidth, height=7.0cm]{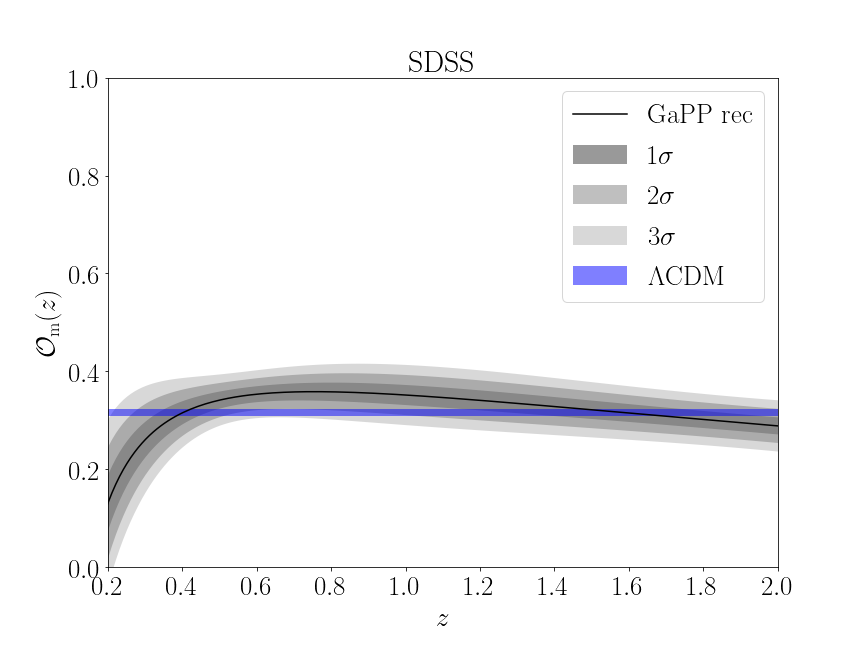}
    \includegraphics[width=0.49\linewidth, height=7.0cm]{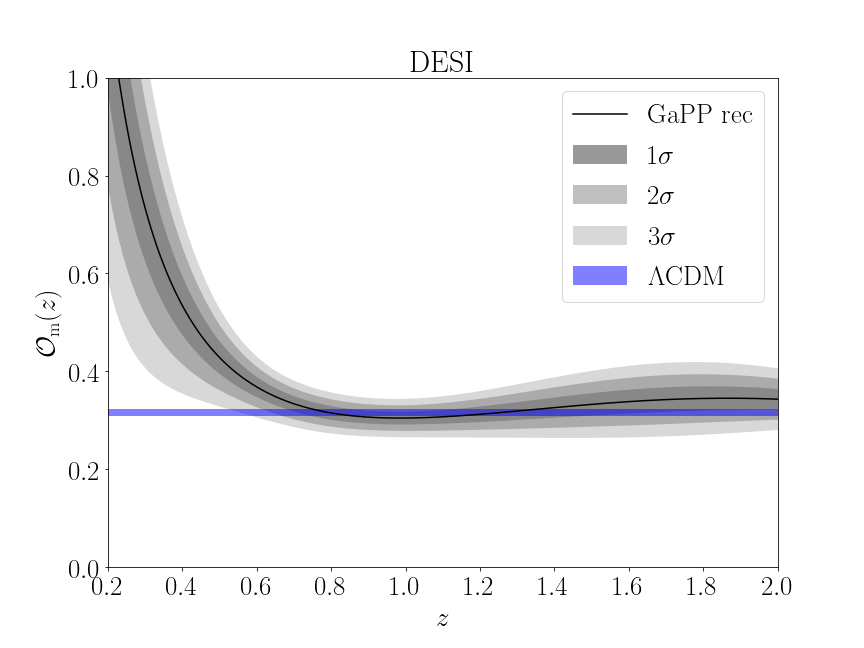}
    \includegraphics[width=0.49\linewidth, height=7.0cm]{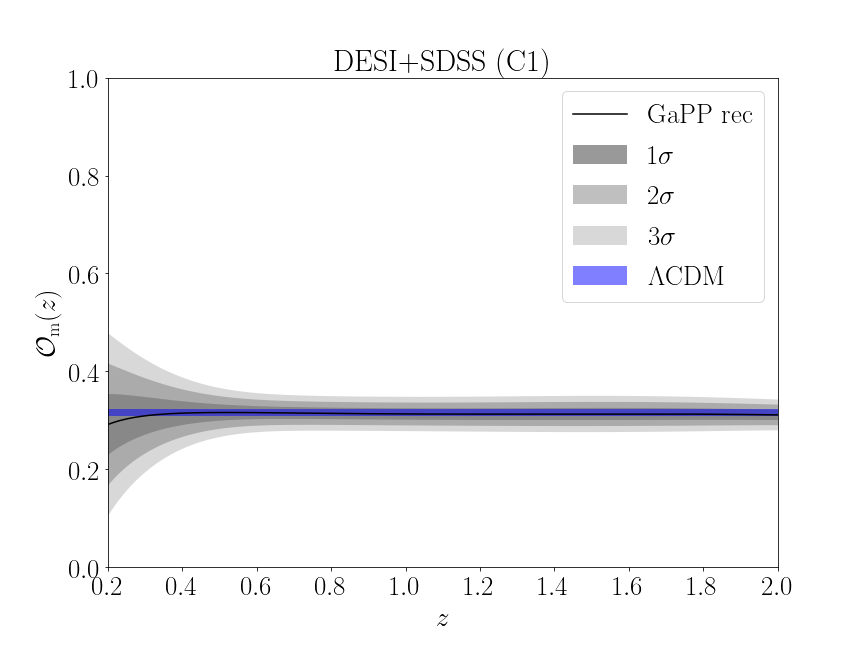}
    \includegraphics[width=0.49\linewidth, height=7.0cm]{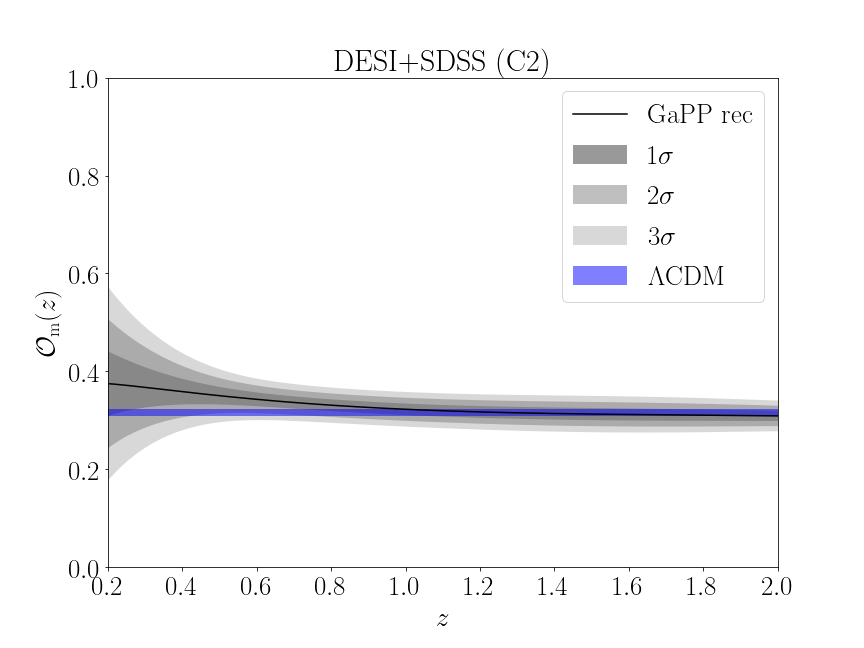}

    \caption{Same as Fig.~\ref{fig:omz_recon}, but assuming the Matern(7/2) GP kernel instead of the squared exponential one.}\label{fig:omz_mat72}
\end{figure}

\begin{figure}[!ht]
    \centering
    \includegraphics[width=0.49\linewidth, height=7.0cm]{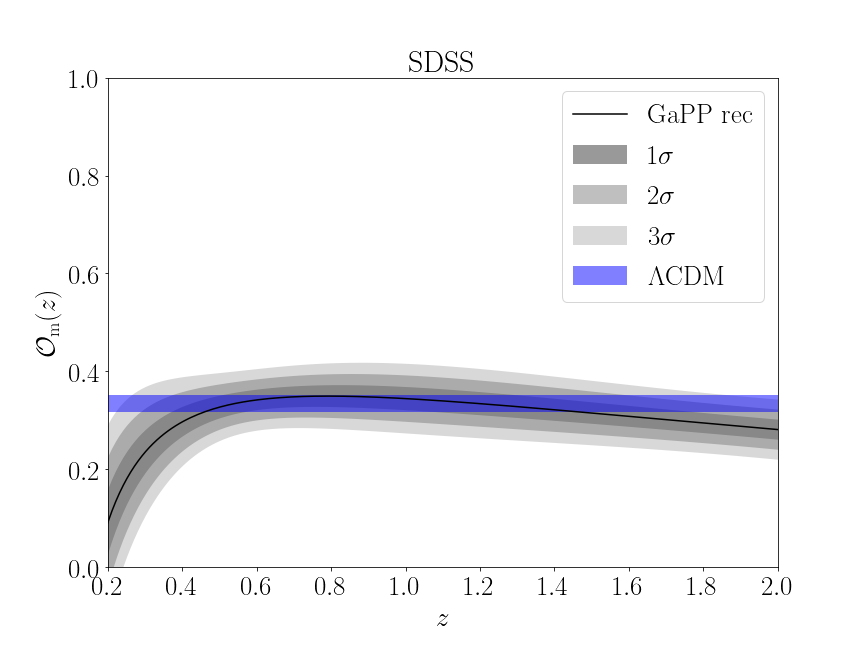}
    \includegraphics[width=0.49\linewidth, height=7.0cm]{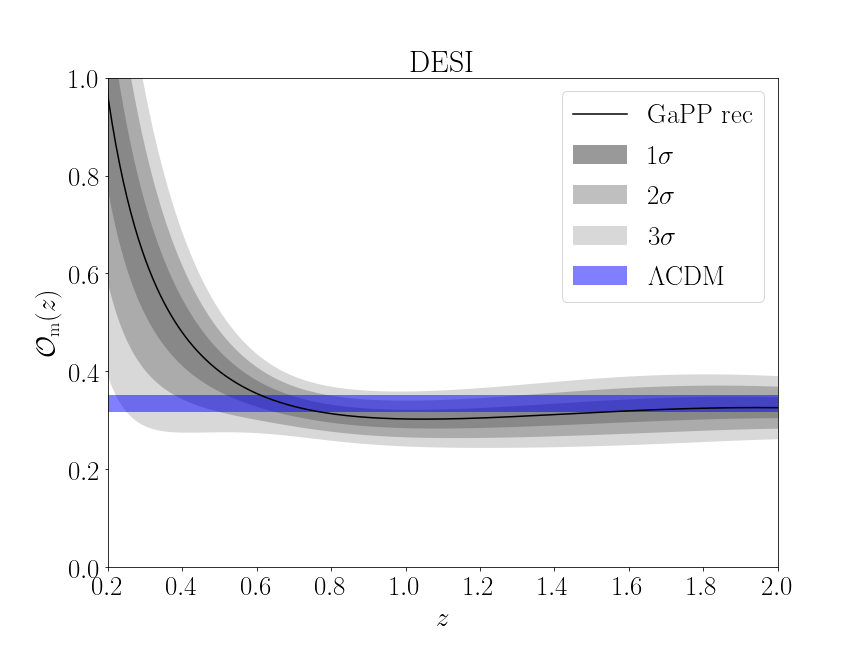}
    \includegraphics[width=0.49\linewidth, height=7.0cm]{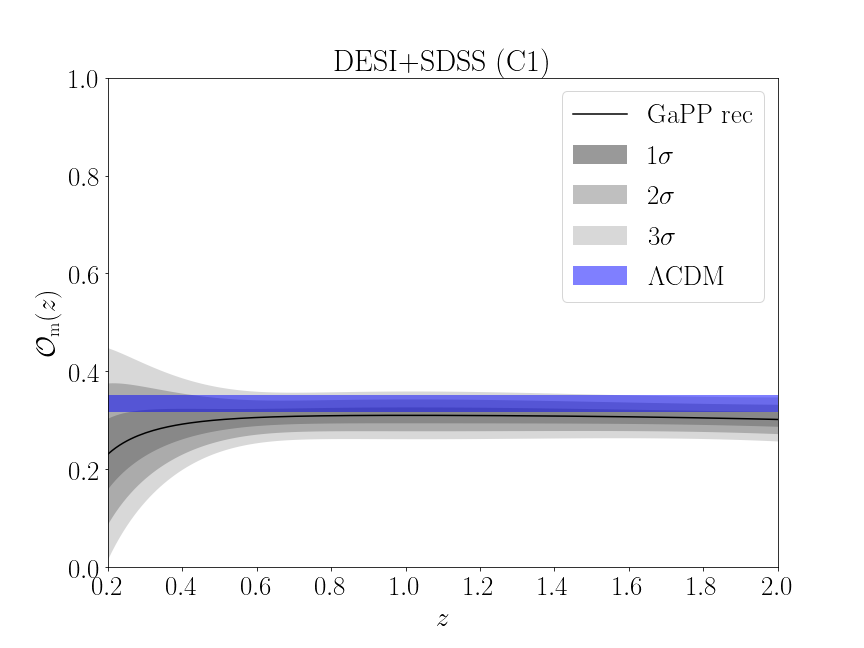}
    \includegraphics[width=0.49\linewidth, height=7.0cm]{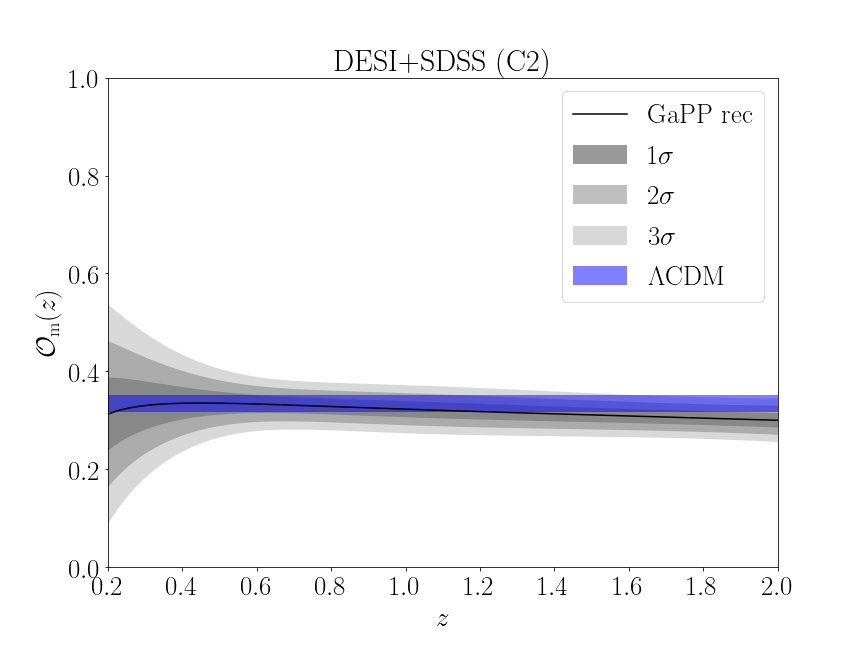}

    \caption{Same as Fig.~\ref{fig:omz_recon}, but assuming the low-$z$ prior on the sound horizon scale $r_{\rm d}$, such as $r_{\rm d} = 136.4 \pm 3.5$ Mpc, as shown in~\cite{Planck:2018vyg}. The thick blue line correspond to a flat $\Lambda$CDM model given by $\Omega_{\rm m} = 0.334 \pm 0.018$, as provided by the PantheonPlus and SH0ES latest results~\cite{Brout:2022vxf}. Correspondingly, we assume a $H_0$ prior by SH0ES in this case, i.e., $H_0 = 73.3 \pm 1.04 \; \mathrm{km} \; \mathrm{s}^{-1} \; \mathrm{Mpc}^{-1}$.}\label{fig:omz_rdlowz}
\end{figure}

Moreover, we check whether these results are robust with respect to different prior assumptions of the reconstructions, namely, other GP kernels, and other values for the sound horizon scale $r_{\rm d}$. The $q(z)$ reconstructions obtained assuming the Matern(7/2) GP kernel are shown in Fig.~\ref{fig:qz_mat72}, whilst Fig.~\ref{fig:qz_rdlowz} presents the $q(z)$ results obtained when we assume $r_{\rm d} = 136.4 \pm 3.5$ Mpc, i.e., if the sound horizon scale measured from low-$z$ cosmological probes, such as SN distances and BAO, as presented in~\cite{Planck:2018vyg}. In the latter case, we remark that the blue dot-dashed lines correspond to a flat $\Lambda$CDM model given by $\Omega_{\rm m} = 0.334$ and $H_0 = 73.3 \; \mathrm{km} \; \mathrm{s}^{-1} \; \mathrm{Mpc}^{-1}$, which is consistent with the latest results from PantheonPlus and SH0ES~\cite{Brout:2022vxf}. We can see that the agreement between the SCM and the DESI/SDSS data alone slightly improves, although we note the same trends of the $q(z)$ curve in both cases. Nonetheless, we find again that the DESI+SDSS C1 and C2 combinations are in better concordance with the SCM prediction than the individual datasets, regardless of the GP kernel and cosmological priors under assumption.

The results of the null diagnostic $\mathcal{O}_{\rm m}(z)$ for both Matern(7/2) kernel and low-$z$ sound horizon values are presented in Figs.~\ref{fig:omz_mat72} and~\ref{fig:omz_rdlowz}, respectively. In the latter case, the thick blue line corresponds to $\Omega_{\rm m} = 0.334 \pm 0.018$, and we assumed the prior $H_0 = 73.3 \pm 1.04 \; \mathrm{km} \; \mathrm{s}^{-1} \; \mathrm{Mpc}^{-1}$ to compute $\mathcal{O}_{\rm m}(z)$ and its uncertainty, i.e., the best-fitted $\Omega_{\rm m}$ and $H_0$ parameters, respectively, assuming a flat $\Lambda$CDM model to the PantheonPlus and SH0ES data~\cite{Brout:2022vxf}. While the alternative GP kernel yields similar results to the default kernel choice, the results are quite different for some specific datasets when we assume the low-$z$ sound horizon measurement and the Hubble constant. In this case, we can still see the same trends for SDSS and DESI datasets alone, albeit with a slightly lower significance. This is also supported by recent works \cite{Wolf:2024eph,Wolf:2024stt}, whereby it is shown that there are no conclusive evidences to consider thawing quintessence models as the description for evolving dark energy. However, we find that the agreement between DESI+SDSS C1 data and SCM marginally worsens when the low-$z$ $r_{\rm d}$ and $H_0$ values are assumed (at slightly over $1\sigma$ confidence level). But even in this specific case, we can still see the same general trend of our main results, that is, the combination of DESI and SDSS data is in better agreement with the SCM predictions than the individual datasets. So, overall, the results obtained from different assumptions on the GP kernel and sound horizon scale present the same trend as in the default case. This demonstrates the robustness of our results with regards to them. Interestingly, however, recent results \cite{Jiang:2024viw} suggest that the upper limits on the neutrino mass sum as obtained from DESI are tightened further if quintessence or non-phantom dark energy models are considered, warranting further investigation into the preference of one class of model over another.

In light of all these results, it thus becomes imperative to understand the caveats in the measurements from these surveys on an individual basis, and predict whether combining these datasets can yield more accurate results.

\section{Conclusion}\label{sec:conc}

The recently released DESI BAO results have sparked a lot of conversation around a possible deviation from $\Lambda$CDM cosmology, hinting at the existence of an evolving dark energy at present times. This is in accordance with parameterised models such as Chevallier-Polarski-Linder (CPL). In this work, we attempt to assess these results in a model-independent manner using Gaussian Processes, whereby we go a step further and also compare the DESI BAO results with that of SDSS BAO. Our results consist of comparing four datasets for each reconstruction: the DESI data and SDSS data taken individually, and two combinations of both - one overall combination, and another in which we combine the high redshift DESI data only ($z>0.7$) with SDSS. We observe that for the non-parametric reconstructions of the Hubble parameter $H(z)$, its derivative $H'(z)$, the deceleration parameter $q(z)$ and the null diagnostic $\mathcal{O}_{\rm m}(z)$, the DESI and SDSS data alone show considerable deviations from $\Lambda$CDM predictions, whereas their combinations are in remarkable agreement with the same. The $q(z)$ reconstruction is particularly intriguing, as here the SDSS data seem to suggest that there has been no accelerated expansion of the universe. However, as expected from the DESI results, we see hints of slower accelerated expansion at present times. A combination of both datasets is in good agreement with $\Lambda$CDM, as mentioned earlier.
\par
Furthermore, the $\mathcal{O}_{\rm m}(z)$ diagnostic seems to suggest that DESI and SDSS alone prefer different behaviour for dark energy between themselves. We further carry out our analysis with a different GP kernel, Matern72, instead of the squared exponential one, as well as impose a different sound horizon scale assuming a low-$z$ prior. Barring some minor changes, the overall behaviour of the reconstructions for all the above cases remains the same. We believe that our analysis calls for further investigation of existing results in order to address this inconsistency between SDSS and DESI. We can expect to have a better insight once the low redshift data of DESI become available and can be compared with SDSS data at similar redshifts. In the meantime, it would be worthwhile to explore these disagreements further so that biased conclusions on the validity of the standard model can be avoided.\\

{\it Acknowledgments:} 
CB acknowledges financial support from Funda\c{c}\~ao \`a Pesquisa do Estado do Rio de Janeiro (FAPERJ) - Postdoc Recente Nota 10 (PDR10) fellowship. BG would like to acknowledge financial support from the DST-INSPIRE Faculty fellowship (Grant no. DST/INSPIRE/04/2020/001534).

\bibliography{paper_v03_clean}

\end{document}